\documentclass[11pt,a4paper]{article}
\usepackage{jheppub}
\usepackage{amsmath}
\DeclareMathOperator*{\argmax}{arg\,max}
\DeclareMathOperator*{\argmin}{arg\,min}
\usepackage{amssymb}
\usepackage{bbold}
\usepackage{bm}
\usepackage{color,graphicx}
\usepackage{slashed}
\usepackage{comment}
\usepackage{hyperref}
\usepackage{subcaption}
\usepackage{makecell}
\usepackage[dvipsnames]{xcolor}
\usepackage{pifont}
\usepackage{caption}
\usepackage{float}
\usepackage{array}
    \newcolumntype{P}[1]{>{\centering\arraybackslash}p{#1}}
    \newcolumntype{M}[1]{>{\centering\arraybackslash}m{#1}}

\preprint{ANL-189807}

\title{Explainable AI classification for parton density theory}

\author[a]{Brandon Kriesten}
\author[a,b]{, Jonathan Gomprecht}
\author[a]{, T.~J.~Hobbs}
\affiliation[a]{High Energy Physics Division, Argonne National Laboratory, Lemont, IL 60439}
\affiliation[b]{Department of Physics, University of Arizona, Tucson, AZ 85721}
\emailAdd{bkriesten@anl.gov}
\emailAdd{tim@anl.gov}

\abstract{
Quantitatively connecting properties of parton distribution functions (PDFs, or parton densities) to
the theoretical assumptions made within the QCD analyses which produce them has been a longstanding problem
in HEP phenomenology.
To confront this challenge, we introduce an ML-based explainability framework, \texttt{XAI4PDF}, to classify PDFs by
parton flavor or underlying theoretical model using ResNet-like neural networks (NNs).
By leveraging the differentiable nature of ResNet models, this approach deploys guided backpropagation to dissect relevant features of
fitted PDFs, identifying $x$-dependent signatures of PDFs important to the ML model classifications.
By applying our framework, we are able to sort PDFs according to the analysis which produced them
while constructing quantitative, human-readable maps locating the $x$ regions most affected by the internal theory
assumptions going into each analysis.
This technique expands the toolkit available to PDF analysis
and adjacent particle phenomenology while pointing to promising generalizations.
}

\date{\today}
\begin{document}
\maketitle

%

\section{Introduction}
\label{sec:intro}
Parton distribution functions (PDFs) are an indispensable input informing
standard model (SM) theory predictions for high-energy experiments.
In the case of ongoing measurements at the Large Hadron Collider (LHC), its planned high-luminosity upgrade
(HL-LHC)~\cite{Apollinari:2017lan,Cepeda:2019klc}, and the upcoming Electron-Ion Collider (EIC)~\cite{Accardi:2012qut},
the collinear unpolarized PDFs of the proton play a central role, given their importance in determining
standard model (SM) baselines for numerous processes while carrying signatures of the still imperfectly understood
dynamics of QCD.
In many instances, control over the PDFs limits measurements at colliders both in their sensitivity to new physics beyond
the standard model (BSM) as well as the statistical rigor of associated precision tests of the
SM~\cite{ZEUS:2019cou,Carrazza:2019sec,Greljo:2021kvv,Madigan:2021uho,Gao:2022srd,Iranipour:2022iak}.
This reality has motivated a worldwide effort~\cite{Alekhin:2017kpj,Barry:2018ort,Hou:2019efy,Bailey:2020ooq,NNPDF:2021njg,PDF4LHCWorkingGroup:2022cjn,Kotz:2023pbu} to model the PDFs in a systematically-improving fashion while
continuously updating extractions in light of a steadily growing set of high-energy data recorded at the LHC
and other facilities.

By their comprehensive nature,
modern PDF fits entail many internal pieces~\cite{Kovarik:2019xvh} --- among them, aspects related to the
deployment of perturbative QCD and electroweak theory; a host of nonperturbative effects essential to the
modeling of highly sensitive data ({\it e.g.}, the inclusion of non-leading twist or nuclear corrections);
the curation and treatment of hadronic data and their accompanying (correlated) systematic and statistical
uncertainties; methodological choices related to analysis procedures and
workflows in practical PDF fits; and open issues in statistical theory which become unavoidable as the
parametric complexity of PDF analysis continues to expand.
Notably, collinear PDF analyses take place within a still larger landscape of quantum correlation functions (QCFs) of
higher dimensionality, including transverse-momentum dependent (TMD) PDFs as well as generalized parton distribution
functions (GPDs)~\cite{Ji:1996ek, Radyushkin:1997ki}; these are central to studies of the nucleon's 3D structure, involving
global data sets including deeply virtual Compton scattering and related experiments planned at Jefferson Lab, the EIC, and
beyond~\cite{Ji:1996nm,Belitsky:2001ns,Kriesten:2019jep,Kriesten:2020wcx,Guo:2022cgq,Accardi:2023chb,Qiu:2022pla,Qiu:2022bpq}.
While there is some success in modeling these functions~\cite{Kriesten:2021sqc,Bertone:2021wib,Guo:2023ahv,Moffat:2023svr}, it
remains an open question in how to perform a global analysis with full consideration of all theoretical and experimental
constraints.
QCD global analyses have progressed to a sufficient level that an era of PDF-driven phenomenology with $\sim$1\% precision is
increasingly plausible~\cite{Amoroso:2022eow} for many high-energy observables of interest.
At the same time, however, increasing working accuracy in PDF fits and the accumulation of high-energy data are such that
variations in analysis assumptions can produce fluctuations in likelihood functions and/or extracted PDFs which had
previously been minor relative to the comparatively larger uncertainties which typified previous-generation analyses.
For instance, improved control over scale uncertainties achieved at NNLO relative to NLO is such that inclusion of
few-percent corrections in other sectors of the global analysis --- for example, related to light nuclear
corrections relevant for key data sets~\cite{Accardi:2021ysh} --- may induce $\chi^2$ variations of similar magnitude to those obtained from shifts in the assumed NNLO formalism. 
As a consequence, it is increasingly necessary to track a significant number of internally coupled degrees-of-freedom in modern
PDF fits, beyond selections of data sets or PDF parametrization forms.

Extracting clear insights regarding how theory or analysis assumptions made in a given
fitting framework influence the behavior of the PDFs they ultimately determine remains the subject of ongoing research.
Historically, the most straightforward approach to assessing the relationship between analysis settings and
fitted PDFs was the construction of arrays of fits in serial fashion. While simple to execute, this method is both computationally
costly and carries the risk of obscuring subtleties owing to the relative smallness of effects involved as well as the inherent
difficulty of separating the combined impacts of simultaneous variations in analysis settings; these latter effects may interfere
both constructively or destructively in surprising ways, potentially conspiring to obscure the effect of varying one
internal parameter against another.
In the meantime, the growing complexity of QCD analyses as noted above greatly heightens the challenge of interpreting
the origin of specific features observed in the fitted PDFs they produce.

To ameliorate this problem, recent efforts have relied on detailed benchmarking of PDF fits, with a strong focus on
achieving a common basis on which to compare global fits as in the
PDF4LHC~\cite{PDF4LHCWorkingGroup:2022cjn,Butterworth:2015oua,Ball:2012wy,Botje:2011sn} exercises.
Similarly, a number of methods based in classical statistical theory have been developed based on metrics to quantify the
statistical pulls on PDFs derived from variations in the
$\chi^{2}$ or benchmark comparisons to higher-level calculated physics processes.
For example, Ref.~\cite{Hobbs:2019gob} proposed the notion of $L_2$ sensitivity, discussed in slightly more detail in
Sec.~\ref{sec:pheno_id} and Ref.~\cite{Jing:2023isu},
which offers a quantitative metric to assess how shifting PDFs within their $1\sigma$ uncertainties induces corresponding
shifts in the $\chi^2_E$ of key experiments constraining the global fit. This method furnishes $x$-dependent information
on the PDF-level pulls of the most sensitive data sets, and builds upon the ($L_1$) PDF Sensitivity, $|S_f|(x,Q)$, introduced in
Ref.~\cite{Wang:2018heo} to quantify and visualize the projected PDF sensitivities of data sets within the context of a
baseline assumed fit.
Both approaches are principally developed within the Hessian global fit method and contributed to the analysis in the most recent
PDF4LHC21 study~\cite{PDF4LHCWorkingGroup:2022cjn}.
Motivated in part by the growing size of global fits and use of neural networks (NNs) in PDF parametrization and related
analysis elements,
a number of recent studies have implemented ML methods, including generative models, to understand properties of PDFs and
their uncertainties; these have included inverse mapper solutions with autoencoders~\cite{9534012,Kriesten:2023uoi}, Gaussian mixture models~\cite{Yan:2024yir}, and GANs~\cite{Carrazza:2021hny}, which have generally explored efficient representations or reconstructions of PDFs as well as studies of their
associated uncertainties.
These approaches offer parallel and complementary avenues to exploring the complicated correlations between PDFs and their internal
analysis elements.
In Ref.~\cite{Kriesten:2023uoi}, for example, inverse problem solvers were
used to reconstruct PDFs from their Mellin-space behavior ({\it i.e.}, their lattice-calculable moments); the resulting interpretable framework
could be used as a generative model to predict PDFs, interpolate among encoded $x$-dependent shapes, or explore connections between PDFs and
higher-dimensional distributions like GPDs.
In this study, the initial input space of (variational) autoencoders was informed by $x$-gridded arrays of
PDFs obtained from toy-model parametrizations. A natural extension of these approaches involves training feed-forward networks
on phenomenological PDFs themselves as in Ref.~\cite{Liu:2022plj}, an approach which allows a variety of fast studies, including
rapid exploration of combined PDF-EFT theory spaces~\cite{Gao:2022srd}.
In addition, some groups have implemented machine learning (ML) methods~\cite{Kumericki:2016ehc,Cuic:2020iwt,Almaeen:2022imx,Almaeen:2024guo} to understand how physics
constraints such as PDF limits can be incorporated into full fits of the GPDs. 
Although these foregoing studies enlarge the toolkit of PDF analysis methods, they have largely focused on the
task of encapsulating the PDFs themselves in NN parametrizations and/or training to a likelihood function, rather than
attempting to disentangle the role model assumptions play in fitted PDFs, let alone doing so in a common setting which
simultaneously admits collections of global fits.
Moreover, previous investigations were more immediately related to tasks of PDF reconstruction or regression; meanwhile, an adjacent
problem is that of performing model discrimination among distinct families of parton densities --- a problem we recognize as inherently
one of classification.
Motivated by this problem as well as the need to diagnose the PDF-level impacts of assumed theory in an apple-to-apples framework,
in the present work we propose a new, generalizable scheme based on the ML modeling of fitted phenomenological PDFs.
Specifically, we concentrate on model discrimination through classification\footnote{In a forthcoming study~\cite{XAIreg}, we revisit the question of regression within an explainable AI framework.} within a closed set of PDF analyses, as this provides a means of cleanly distinguishing among
unique phenomenological fits (when statistically possible) and applying explainability methods to trace the ability of ML models to identify
phenomenological fits based on unique features of the predicted PDFs.
This technique, which we call \texttt{XAI4PDF}, enables us in turn to dissect the dependence of the fitted PDFs on elements of the assumed theory
by leveraging recent developments in {\it explainable} AI (XAI).
In addition, by structuring our model around the logic of QCD global fits and thereby integrating domain knowledge, the resulting
approach helps bridge the gap between ML-driven modeling and traditional methods for QCD analysis.

Using a ResNet-like NN architecture optimized for classification, we identify PDFs from the CTEQ-TEA (CT) family of global fits --- specifically,
the most recent NNLO main release, CT18~\cite{Hou:2019efy}. For our proof-of-principle classification tasks, we concentrate on sorting parton
densities according to their associated parton flavor and/or the specific set of global analysis assumptions which produced them. Saliency maps are constructed using gradients calculated by a modified autodifferentiation algorithm called \textit{guided backpropagation}. These saliency maps trace the impact of theoretical assumptions and the contribution of significant experiments to the $x$ and flavor dependence of the PDFs.
The procedure described here also introduces a notion of analysis simultaneity --- rather than serially comparing
one pair of analysis settings and corresponding fitted PDFs, it becomes possible to train a space
of PDF analyses of essentially arbitrary size while classifying the results within a consistent (ML)
model or framework.

Our paper is organized as follows: in Sec.~\ref{sec:theory}, we discuss
the phenomenological and theoretical background that motivate this work, particularly related to PDFs. 
We explain the guided backpropagation technique as related to classification algorithms in Sec.~\ref{sec:XAI}.
In Sec.~\ref{sec:class_model} we describe the model architecture and PDF data that we use to train.
Sec.~\ref{sec:framework} is where we provide details of our application of the \texttt{XAI4PDF} framework
in two example cases for PDFs: in Sec.~\ref{sec:pdf_flavor_id} we use classifying CT18 NNLO PDF replicas by
flavor, and in Sec.~\ref{sec:pheno_id} we classify PDF replicas by phenomenological fit with underlying theory
assumptions. Finally, we summarize conclusions and possible future directions in Sec.~\ref{sec:conc}.
%
%
%
\section{Parton densities: background theory and phenomenology}
\label{sec:theory}
PDFs are inherently probabilistic objects providing the likelihood of finding a parton of flavor, $f$, carrying some fractional
momentum of the proton, $x$, at a given hard scale, $Q$.
BSM searches now engage a variety of channels, such as inclusive and coincident Higgs production;
processes involving $W^{\pm}$/$Z$ gauge bosons~\cite{ATLAS:2016nqi}; and various top-production~\cite{Ablat:2023tiy}
channels. In many cases, these experiments not only provide fertile BSM search contexts, but
also constraining data on hadronic structure which may be actively included in PDF global analyses. 
Predictions for collider processes are made on the basis of QCD factorization theorems~\cite{Collins:1985ue,Collins:2011zzd,CTEQ:1993hwr}, which render the
PDF dependence of theoretical cross sections explicit.
In a simple but ubiquitous example, the single-inclusive production of an electroweak boson in $pp$ scattering at the LHC,
$p_1 + p_2 \to W^\pm/Z + X$, is given schematically by the convolution of a short-distance matrix element, $\hat{\sigma}_{ij}$, with
unpolarized proton PDFs as
\begin{equation}
\label{eq:fact}
{\sigma^{pp \to W/Z} \over dy_{W/Z} }\ =\ \sum_{i,j} \int dx_1 dx_2\, f_i(x_1) f_j(x_2)\, {d \hat{\sigma}^{W/Z}_{ij} \over dy_{W/Z}} (x_1, x_2)\ .
\end{equation}
The parton-level arguments entering hard matrix elements and PDFs in the expression above are typically evaluated
via Born-level kinematical matchings, such that, for the gauge-boson production scenario above,
the relevant parton fractions and hard scales are
\begin{equation}
x_1, x_2 \sim {M_{W/Z} \over \sqrt{s}}\, \exp {\pm y_{W/Z}}\ , \ \ \ \ \ Q \sim M_{W/Z}\ ,
\label{eq:kinematics}
\end{equation}
respectively. We note that this matching is subject to higher-order corrections motivating the inclusion of greater theoretical
working accuracy in both $\alpha_s$ and the electroweak interaction strength. In particular,
from Eq.~(\ref{eq:kinematics}) we conclude that rapidity-dependent cross sections,
$d\sigma / dy_{W/Z}$, involve an $x$ dependence spanning well beyond the nominally expected
low-$x$ region, with forward-boosted electroweak events probing the high-$x$ behavior of the PDFs,
in addition to possible suppressed BSM contributions.
This reality underscores the fundamental challenge of PDF phenomenology for high-stakes HEP observables:
precision control for quantities like EW boson rapidity or invariant-mass distributions requires
detailed knowledge of the PDFs over a wide range of $x$ and a multiplicity of parton-level channels, with combinations
like $\sim\! \bar{u}(x_1) u(x_2), \dots, \bar{s}(x_1) s(x_2)$ relevant for $Z$, or
$\sim\! \bar{d}(x_1) u(x_2), \dots, \bar{s}(x_1) c(x_2)$ implicated in $W^+$ production.
Moreover, these PDFs must be DGLAP-evolved~\cite{Gribov:1972ri,Dokshitzer:1977sg,Altarelli:1977zs} to the appropriate
hard scale, $Q$, for a given experiment, a process which itself introduces complicated entanglements among the
PDFs' dependence on flavor, $x$, and $Q$.
These challenges are such that the only currently-available path to determine PDFs for precision HEP phenomenology
is performing comprehensive QCD global analyses of the world's hadronic data.

Phenomenological PDF analyses have an established history dating to the early 1990s, with a
variety of methods blossoming in the ensuing decades.
The logic of QCD global analysis is commonly shared across these approaches as motivated by the above discussion, with parameters fitted using
data from a diverse set of hadronic processes, including deep inelastic scattering (DIS), Drell-Yan (dilepton) production, top-quark processes,
and $W^{\pm}$/$Z$ gauge-boson production as noted around Eq.~(\ref{eq:fact}). State-of-the-art implementations of perturbative calculations are currently at the order of NNLO in the strong coupling, $\alpha_{s}$, although
there is ongoing work on incorporating specific analysis elements at N$^{3}$LO to produce approximate
aN$^{3}$LO fits~\cite{McGowan:2022nag,Cooper-Sarkar:2024crx,NNPDF:2024nan}; there is not yet sufficient progress, however,
to obtain a comprehensive fit at true N$^{3}$LO.
Additionally, there are many nuanced theoretical considerations for a full phenomenological fit which we will not address in this study.

\begin{figure}[!ht]
    \centering
    \includegraphics[width=\columnwidth]{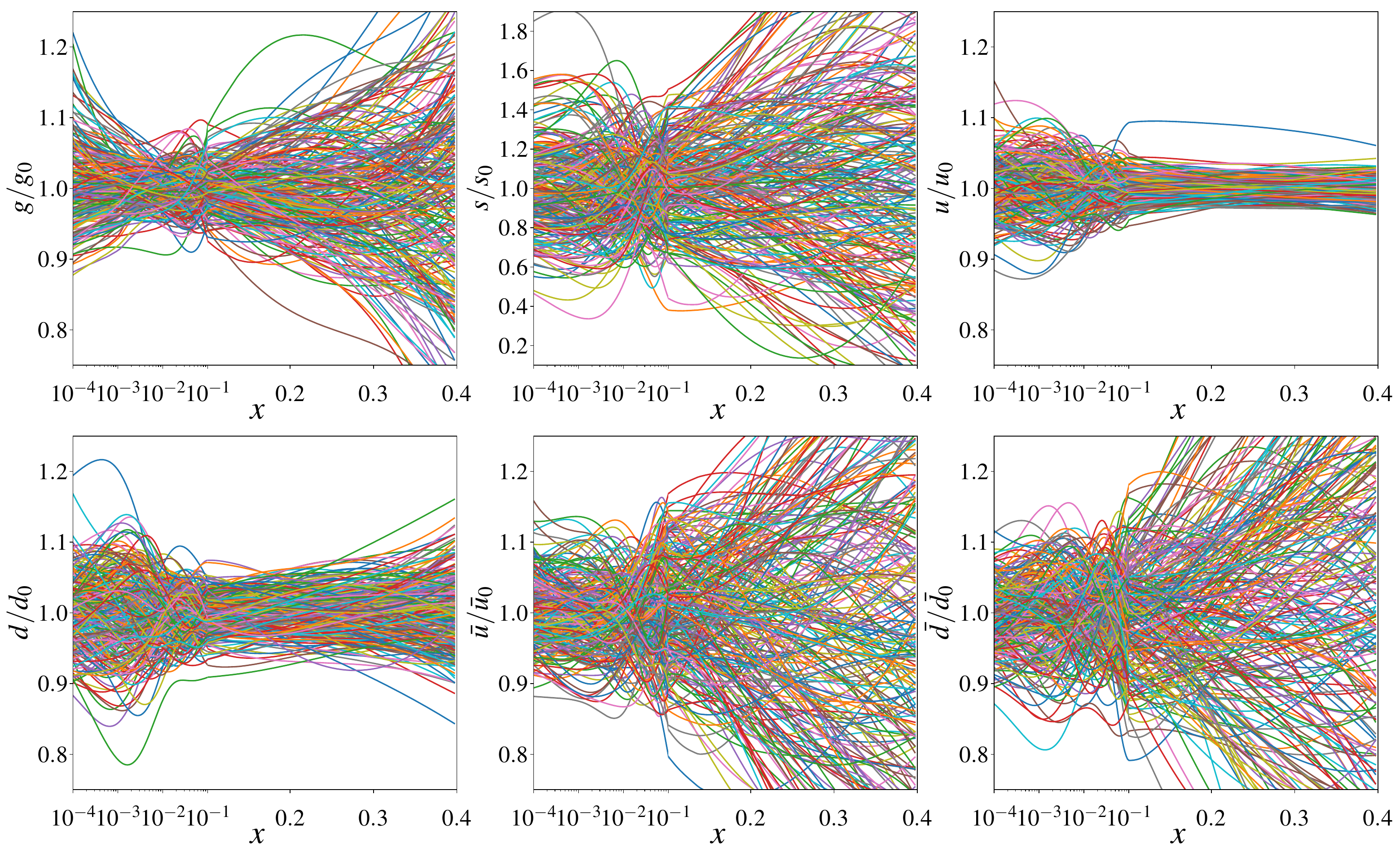}
    \caption{Monte Carlo replicas generated for the CT18 NNLO fit, plotted as ratios to the central-fitted PDF; we take a perturbative scale $Q^{2} = 100$ GeV$^{2}$ and display the range $x = [10^{-4} - 0.4]$ over which the ResNet model inputs (training data) are sampled.}
    \label{fig:pdf_data}
\end{figure}

Ultimately, despite the considerable variation in analysis settings implied above by the wide range of data sets and theoretical
choices available to QCD fits, the great preponderance of global analyses falls into one of two main categories: those involving
explicit analytical (functional) forms to parametrize the $x$ dependence of the PDFs, and methods involving (feed-forward) neural
networks (NNs)~\cite{NNPDF:2021njg} as the basis for PDF parametrization.
Effectively, these parametrizations assume a common form partly motivated by Regge theory, with flavor-specific
PDF combinations given by
\begin{equation}
\label{eq:param}
f_i(x, Q) = N_i x^{\alpha_i}\, (1-x)^{\beta_i}\ \mathcal{P}_{i}(x)\ ,
\end{equation}
and higher parametric flexibility attained by taking a polynomial of arbitrary rank, $\mathcal{P}_{i}(x)\! =\! P^{j}_i(x)$,
in the analytic case~\cite{Kotz:2023pbu}, or a NN, $\mathcal{P}_{i}(x)\! =\! \mathrm{NN}_i(x)$~\cite{NNPDF:2021njg}, in which the PDF shapes are
encoded by the activation functions and weights of the network.
Alongside a chosen parametrization basis, global analyses must also adopt a prescription for the comprehensive determination of the
associated PDF uncertainties. In this respect, the commonly used paradigms are the Hessian formalism~\cite{Hou:2019efy,Bailey:2020ooq,Alekhin:2017kpj} typically employed alongside analytical PDF parametrizations, and the
method of Monte Carlo (MC) resampling~\cite{Costantini:2024wby}. In the former approach, a Hessian matrix is diagonalized at the global minimum
of the analysis likelihood function, $\chi^2$, allowing a basis of eigendirections to be defined; with these, it is then possible to saturate
a tolerance criterion through finite displacements along each eigendirection, such that the parametric dependence of the likelihood
function can be enfolded into the uncertainty via a finite collection of PDF eigensets. The approach of MC resampling
is commonly used in both NN parametrizations of the PDFs~\cite{NNPDF:2021njg} as well as fits based on analytical
forms~\cite{Hunt-Smith:2022ugn}; in this scenario, statistical ensembles of PDF error replicas are generated by optimizing model parameters to a set
of pseudodata sampled from Gaussian distributions about the fitted data in the analysis. The resulting MC replicas then collectively encapsulate the PDF uncertainty.
In the present study, given the need to generate sufficiently plentiful instances of PDF behavior to train
the ResNet-like model introduced in Sec.~\ref{sec:intro}, we assume the validity of MC resampling for the purpose
of quantifying the true uncertainty, following the examples of Refs.~\cite{Kriesten:2023uoi,Liu:2022plj},
and generate sizable sets of replicas as shown in Fig.~\ref{fig:pdf_data}.

This is possible due to the fact that systematic conversions between the Hessian and MC approaches have been developed~\cite{Hou:2016sho}.
For this work, PDF MC replicas, labeled by index $(k)$ in the following equations, are created from the Hessian error sets using the \texttt{mcgen} code~\cite{Hou:2016sho} wherein the central fitted PDF, $X_{0}$, is randomly displaced by a replica-specific value, $d^{(k)}$, and global shift, $\Delta$:
\begin{eqnarray}
    X^{(k)} = X_{0} + d^{(k)} - \Delta \ .
\end{eqnarray}
The random displacements are given by positive and negative variations along successive Hessian eigendirections, $X_{+i}$ and $X_{-i}$, about the central-fitted PDF, $X_{0}$, according to a random Gaussian sampling, $R_{i}^{(k)}\! \sim\! \mathcal{N}(0,1)_{i}^{(k)}$; these are finally summed over each eigendirection, $i$, among the full set of $D$ eigendirections in the phenomenological fit:
\begin{eqnarray}
    d^{(k)} = \sum_{i=1}^{D} &\Bigg( {1 \over 2}[X_{+i} - X_{-i}]\, R_{i}^{(k)} + {1 \over 2}[X_{+i} + X_{-i} - 2X_{0}]\, \left(R_{i}^{(k)}\right)^{2} \Bigg)\ .
\end{eqnarray}
The global shift is the expectation value of the random displacements,
\begin{eqnarray}
    \Delta &=& \frac{1}{N}\sum_{k=1}^{N}d^{(k)}.
\end{eqnarray}

In the case of the MC replica method, the results of a PDF analysis are encapsulated by the resulting statistical ensemble, with
the mean over the full set corresponding to the optimal fit\footnote{In practice, small deviations can exist between the set of
PDFs corresponding to the global minimum of $\chi^2$ and the statistical mean over the MC ensemble; in the present work,
we explicitly ensure that any such deviations are zero.}; the PDF uncertainties can then be calculated as statistical deviations from
this mean.
In particular, the average (central) prediction in the MC method for a PDF-dependent quantity, $X$,
is denoted as $\langle X \rangle$ and given by the discrete expectation over replicas; namely,
\begin{eqnarray}
    \langle X \rangle \approx X_{0} &=& \frac{1}{N}\sum_{k=1}^{N}X^{(k)}\ .
\end{eqnarray}
While the symmetric PDF uncertainty for $X$ can then be obtained as the standard deviation
of replica-by-replica differences from this mean, the more general case of the asymmetric uncertainty may be determined
from the ensemble of MC replicas at the $68\%$ confidence level (c.l.) as
\begin{eqnarray}
    \delta^{MC>}X &=& + \sqrt{\langle (X - \langle X \rangle)^{2}\rangle_{X > \langle X \rangle} }\nonumber \\
    \delta^{MC<}X &=& - \sqrt{\langle (X - \langle X \rangle)^{2}\rangle_{X < \langle X \rangle} }\ ,
\label{eq:MCshifts}
\end{eqnarray}
where $>$ denotes the $68\%$ c.l.~upward shift from the mean value and $<$ denotes the corresponding shift below. We
note that Eq.~(\ref{eq:MCshifts}) is applied by only admitting those replicas predicting values of
$X$ greater than $\langle X \rangle$ in the calculation of the upward variation, and similarly in the calculation
of the downward asymmetric uncertainty.
Taking $X = f_i(x,Q^2)$, this procedure may be applied on a point-by-point basis for a given value of $x$ for each PDF flavor
to obtain asymmetric PDF uncertainty bands about the mean (central) PDF.
We apply this procedure to generate $10^4$-member MC replica sets for each of the CT18 fits explored in this work;
we show an example of central-set-normalized PDF ratio replicas for CT18 NNLO in Fig.~\ref{fig:pdf_data}.

\begin{table*}[t]  
\scalebox{0.78}{
\begin{tabular}{ccccc}
\hline 
\textbf{PDF} & \textbf{Factorization scale } & \textbf{ATLAS 7 TeV $W/Z$} & \textbf{CDHSW $F_{2}^{p,d}$ } & \textbf{Pole charm }\tabularnewline
\textbf{fits}
 & \textbf{in DIS} & \textbf{data included?\quad} & \textbf{data included?\quad} & \textbf{mass, GeV}\tabularnewline
\hline 
\hline 
\noalign{\vskip6pt}
CT18 NNLO & $\mu_{F,DIS}^{2}=Q^{2}$ & No & Yes & 1.3\tabularnewline[6pt]
\hline 
\noalign{\vskip6pt}
CT18A NNLO& $\mu_{F,DIS}^{2}=Q^{2}$ & Yes & Yes & 1.3\tabularnewline[6pt]
\hline 
\noalign{\vskip6pt}
CT18X NNLO& $\mu_{F,DIS}^{2}=0.8^{2}\left(Q^{2}+\frac{0.3\mbox{ GeV}^{2}}{x_B^{0.3}}\right)$ & No & Yes & 1.3\tabularnewline[6pt]
\hline 
\noalign{\vskip6pt}
CT18Z NNLO & $\mu_{F,DIS}^{2}=0.8^{2}\left(Q^{2}+\frac{0.3\mbox{ GeV}^{2}}{x_B^{0.3}}\right)$ & Yes & No & 1.4\tabularnewline[6pt]
\hline 
\end{tabular}
}
        \caption{
                The four PDF fits analyzed in this work and the differences in their theory assumptions. Table obtained from \cite{Hou:2019efy}.
        }
\label{tab:AXZ}
\end{table*}

To train our classifier network on a theoretically meaningful and self-contained discrimination task, we take the set of global fits accompanying the CT18 NNLO main release~\cite{Hou:2019efy}. This collection consisted of four QCD analyses spanning several variations in the chosen theory and/or global data set: CT18 as well as the three CT18A/X/Z alternative fits, where the different analysis assumptions informing this collection are summarized in Table~\ref{tab:AXZ}. Taken together, the four NNLO fits of CT18 represent the most recent global fit after the CT14~\cite{Dulat:2015mca} and CT14HERA2~\cite{Hou:2016nqm} releases, following the inclusion of a variety of data from Run-1 of the LHC, such as inclusive jet production, $W^{\pm}$/$Z$ boson production, and $t\bar{t}$ pair production from the ATLAS, CMS, and LHCb. Among these,
the CT18 NNLO main fit constitutes a baseline fitting scenario to which the CT18A/X/Z alternative fits were comparative variations allowing
exploration of the relevant physics.
CT18A included an additional data set on 7 TeV $W^{\pm}$/$Z$ production from ATLAS, discussed in detail in App.~A of Ref.~\cite{Hou:2019efy}. Meanwhile, CT18X implemented a modified scaling prescription for low-$x$ DIS, mimicking the effects of low-$x$ resummation. Lastly, CT18Z combined the assumptions of the CT18A and CT18X NNLO fits while also removing the CDHSW $F_{2}^{p,d}$ data and shifting the charm mass to $m_c \!=\! 1.4$ GeV to agree with HERA inclusive DIS and LHC vector boson measurements~\cite{Hou:2019efy}. The CT18 and CT18Z fits are the most distinct from each other in theory space due to their differing underlying theory assumptions.
We stress that the 4-member collection of fit variations provided by CT18 provides a ideal testbed to demonstrate the XAI methods
developed in Sec.~\ref{sec:framework}; this owes to its self-contained nature of this simple ensemble based on a discrete set of analysis
variations; these features collectively make for a training task well-disposed to straightforward ML labeling and classification.

Our data set for training is composed of MC replicas in the 6-fold flavor basis of the CT18 fits, spanned by the $g$-, $s$-, $u$-, $d$-, $\bar{u}$-, and $\bar{d}$-PDFs; given its proximity to the electroweak and Higgs scales, we take each of these at
a common physical scale of $Q^{2}\! =\! 100$ GeV$^{2}$, where $Q = \mu_F = \mu_R$. In practice, we generate MC replicas from the available LHAPDF grids~\cite{Buckley_2015} associated with the original Hessian error sets using default settings in the \texttt{mcgen} package~\cite{Hou:2016sho}. We use a split $x$-scale that is logarithmic in the low-$x$ region, $(10^{-4}\! -\! 10^{-1})$ and linear from $0.1\! -\! 0.4$, with approximately $100$ points in each range for a total of $200$ combined points covering the sampled range in $x$. 
We note that the large- and small-$x$ PDFs are comparatively difficult to constrain experimentally, leading to a sparsity of sensitive data relevant to those regions. Consequently, the PDF shapes at high and low $x$ are significantly determined by the assumed parametrization, with corresponding increases in the PDF uncertainty. To mitigate the impact of the parametrization on the classifications made by the NNs, we elect in this study to train on self-normalized ratios, dividing replicas by their corresponding central PDFs ($f$/$f_{0}$) to produce the information in Fig.~\ref{fig:pdf_data}.

As argued in Sec.~\ref{sec:intro}, it would be beneficial to comprehensively connect PDF analysis theory settings to the $x$-dependence of the associated
PDFs in a common and generalizable framework. We perform this task by classifying PDFs across the CT18 series of phenomenological fits using the
XAI approach discussed in Sec.~\ref{sec:XAI} below.

%
\section{Explainable AI for PDF classification} 
\label{sec:XAI}
Explainable AI (XAI) refers to a collection of techniques used to interpret a ML model's predictions in a human-readable format (see Refs.~\cite{gohel2021explainable, xaioverview21} and references therein). Typically, these techniques come in the form of \textit{saliency maps} which highlight relevant areas of the input feature space that were crucial during the prediction process. The attribution scores calculated with XAI methods come in many forms such as gradients, classification scores, or other numerical ratings. Given our desire to trace the effects of theory assumptions on the $x$-dependent behavior of individual instances of the fitted PDFs, in this study we principally explore {\it local} methods; these explain the decisions of a NN with respect to an individual sample of data. This is contrasted to {\it global} attribution methods which explain decisions of a NN using features which were significant across all inputs. Given the size and complexity of most NN models, global methods are often not feasible. 

Saliency maps are often used in computer vision to define regions where pixels are substantially different from their surroundings. In classification tasks, they highlight the regions of an image which most significantly impact the classification score. We can translate this concept from computer vision to QCD analysis to disentangle important physics that is learned by the NN during classification. There are many different types of saliency methods that use gradients~\cite{simonyan2014deep,sundararajan2017axiomatic,shrikumar2017just, springenberg2015striving, simonyan2014deep, smilkov2017smoothgrad}, class-activated maps~\cite{Selvaraju_2019}, occlusion~\cite{zeiler2013visualizing}, and even edge detectors/image segmentation~\cite{minaee2020image, 4767851}. In this work, we focus on a particular gradient-based saliency map called guided backpropagation~\cite{simonyan2014deep}.

We start with a training data set denoted as $\mathcal{D} = \{\mathbf{x}^{(i)} , y^{(i)} \}$ which contains a set of input features, $\mathbf{x}^{(i)}$ (here, $x$-dependent values of the PDFs), and associated categorical labels, $y^{(i)}$ (corresponding parton flavors or PDF models). Samples from the data set are drawn from the joint probability distribution $p(\mathbf{x},y)$ where $\mathbf{x}$ and $y$ are random variables of the input and label spaces, respectively. For a classification problem with $C$ total classes, the classifier model can be functionally described by the mapping $f(\mathbf{x}, \theta):  \mathbb{R}^{\mathcal{D}} \rightarrow \mathbb{R}^{C}$ through model parameters, $\theta$. The classification model outputs parameters of a categorical distribution represented as a vector of class probabilities, $ \hat{\pi} = \sigma(f(\mathbf{x},\theta))$ where the ML model learns to approximate a categorical multinomial distribution:
\begin{eqnarray}
    \text{Cat}(y|\mathbf{\hat{\pi}}) &=& p(y= c | \mathbf{x}^{(i)}, \theta) \nonumber \\
    &=& \prod_{c=1}^{C} \pi_{c}^{\mathbb{1}(y^{(i)}=c)}\ .
\end{eqnarray}
Here, $\pi_{c}$ represent the model predictions for the probability that the input belongs to the particular class $c$, and the function $\mathbb{1}(\cdot)$ is $1$ when the condition is satisfied and $0$ otherwise.
The index $i$ runs through the full collection of training examples, of number $N$, of feature/label pairs in the data set $\mathcal{D}$.
The set of predicted output scores is numerically unbounded, $\{ f_{c}(\mathbf{x}^{(i)}, \theta) = S_{c}^{(i)}\}_{c=1}^{C} = S^{(i)}$; therefore, in order to predict probabilities, the outputs of this function are passed through a SoftMax activation, $\sigma$, defined as:
\begin{eqnarray}
\label{eq:soft}
    \sigma\big[S^{(i)}\big] = \left[ \frac{e^{S_{c=1}^{(i)}}}{\sum_{c=1}^{C} e^{S_{c}^{(i)}}}, \dots , \frac{e^{S_{c=C}^{(i)}}}{\sum_{c=1}^{C} e^{S_{c}^{(i)}}}\right]\ .
\end{eqnarray}
We train the model through maximum likelihood estimation which can be written in terms of predicted labels, $\displaystyle \hat{y} = \argmax_{c}(\hat{\pi})$, as:
\begin{eqnarray}
    \theta^{*} = \argmax_{\theta}\left(\prod_{i=1}^{N} \prod_{c=1}^{C} p(\hat{y} = c| \mathbf{x}^{(i)}, \theta)^{\mathbb{1}(y^{(i)}=c)}\right)\ .
\end{eqnarray}
It is equivalent, and computationally more tractable, to minimize the negative log likelihood, as doing so stabilizes the training regimen:
\begin{align}
    \theta^{*} =  \argmin_{\theta}\Big[ &\mathcal{L}(\theta) \Big]\ \nonumber \\
    &\mathcal{L}(\theta) =  -\frac{1}{N}\sum_{i=1}^{N} \sum_{c=1}^{C} {\mathbb{1}(y^{(i)}=c)}\, \ln{\left(p(\hat{y}=c | \mathbf{x}^{(i)}, \theta)\right)}
\label{eq:loss}
\end{align}
The latter expression above specifies the categorical cross-entropy loss, and is most often used while training multi-class classification problems.

To further illustrate the mathematical structure of classifier models and coincident use of backpropagation, we consider the functional form of a shallow NN:
\begin{align}\label{fout}
f^{\text{out}} = \sigma^{\ell}\Big[\omega^{\ell}\Big(\sigma^{\ell-1}\big(\omega^{\ell-1}\left(\dots \sigma^{0}\left(\omega^{0} x + b^{0}\right) + \dots \right) + b^{\ell-1} \big) + b^{\ell} \Big)\Big]\ , & 
\end{align}
where $b^{\ell}$ and $\omega^{\ell}$ are the respective biases and weights of layer $\ell$; $\sigma^{0}$, $\dots$, $\sigma^{\ell-1}$ represent ReLU activation functions,
$\text{ReLU}(x) = \text{max}(0,x)$, and $\sigma^{\ell}$ is the SoftMax activation function defined in Eq.~(\ref{eq:soft}) above.
During training, we use autodifferentiation to efficiently compute gradients of the (categorical cross-entropy) loss function, $\mathcal{L}$ of Eq.~(\ref{eq:loss}), with respect to the NN's parameters ({\it i.e.}, its weights, $\omega$, and biases, $b$)~\cite{nielsen2015neural}. 
Schematically, we can write these gradients in terms of a specific layer $n$, where $n$ exists within a larger set of layers, $\{ \ell, \ell - 1, \dots, n+1, n, n-1, \dots, 1, 0\}$:
\begin{eqnarray}
    \frac{\partial \mathcal{L}}{\partial b^{n}} &=& [(\omega^{n + 1})^{T}\delta^{n + 1}]\odot \frac{\partial \sigma}{\partial z}(z^{n}) \label{eq:grad1}\\
    \frac{\partial \mathcal{L}}{\partial \omega^{n}} &=& f^{n-1}[(\omega^{n + 1})^{T}\delta^{n + 1}]\odot \frac{\partial \sigma}{\partial z}(z^{n}) \label{eq:grad2}\ ,
\end{eqnarray}
where $f^{n} = \sigma^{n}(z^{n})$ are the outputs of the $n$th layer, $z^{n} = \omega^{n}f^{n-1} + b^{n}$, and the errors, $\delta$, are backpropagated through the full NN from the final layer, $\ell$, to the initial layer:
\begin{eqnarray}
\label{eq:delta_errs1}
    \delta^{\ell} &=& \nabla_{f}\mathcal{L}\odot \frac{\partial \sigma}{\partial z}(z^{\ell})\\
    \delta^{n} &=& [(\omega^{n + 1})^{T}\delta^{n + 1}]\odot \frac{\partial \sigma}{\partial z}(z^{n})\ .
    \label{eq:delta_errs2}
\end{eqnarray}
In the above notation, the symbol $\odot$ is the Hadamard product which is the element-wise multiplication of two matrices. The backpropagation algorithm begins with a forward pass from input feature space to the calculation of $f^{out}$. At the final layer, $\ell$, the error vector $\delta^{\ell}$ of Eq.~(\ref{eq:delta_errs1}) is computed and propagated back through to the previous layers using Eq.~(\ref{eq:delta_errs2}). The gradients can then be subsequently calculated at each layer using Eqs.~(\ref{eq:grad1}) and (\ref{eq:grad2}). The optimizer uses the gradients to update the network weights and biases. Autodifferentiation automates the application of the chain rule through the functional mapping, tracking the mathematical operations and their derivatives during the forward pass at each layer to facilitate efficient gradient computation during the backward pass.

Once the model is trained, we can use autodifferentiation to compute the gradients of the trained model's pre-SoftMax classification score, $S_{c}^{(i)}$, with respect to the input feature space in a single backward pass \cite{Lerma_2023}.
The guided backpropagation algorithm modifies how this differentiation is performed over the rectified linear unit (ReLU) activations by applying a double filter such that only positive gradients in the current layer and positive inputs from the previous layer are propagated during a single backward pass.
This double masking is accomplished by replacing the ReLU activation layers with custom layers which satisfy the following condition:
\begin{equation}
\frac{\partial f^{\text{out}}}{\partial f_i^{n}} = \left(f_i^n > 0\right) \cdot \left(\frac{\partial f^{\text{out}}}{\partial f_i^{n+1}} > 0\right) \cdot \frac{\partial f^{\text{out}}}{\partial f_i^{n+1}}\ .
\end{equation}
Here, $f_i^n$ refer to the outputs of the $n^\mathit{th}$ layer and $f^{\text{out}}$ is the output of the final pre-SoftMax layer. 
By double masking, we constrain the autodifferentiation procedure to propagate only those gradients that positively influence the classification score. 
This approach ensures that we capture only the essential information necessary for the model's prediction, resulting in precision maps capable of fine detail.

%
\section{ML model architecture and optimization}
\label{sec:class_model}

\begin{figure}
\centering
\includegraphics[width=0.65\columnwidth]{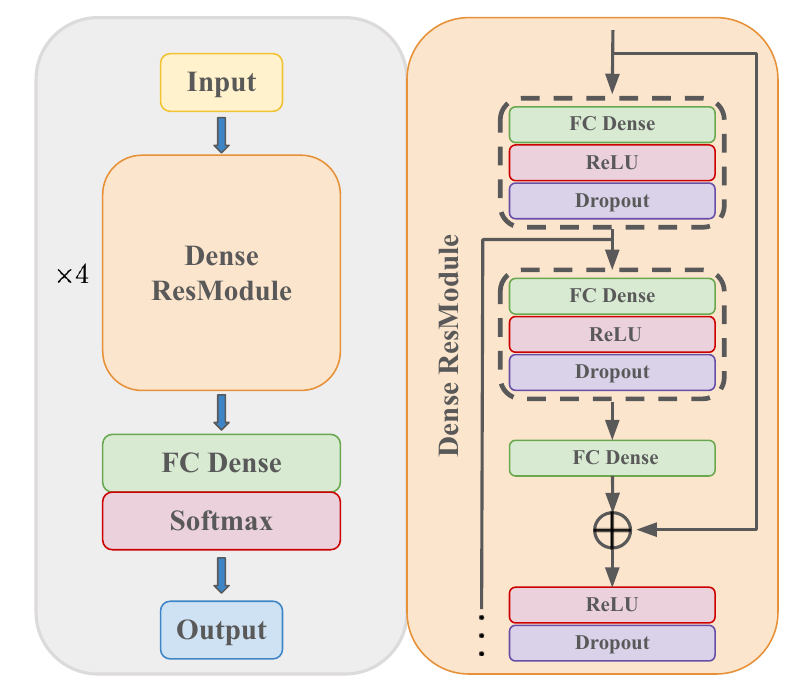}
\caption{The architecture of the ResNet-like NN used to classify PDFs. The model contains an input layer, four dense ResModules, followed by a final fully-connected (FC) dense layer to predict classification scores, $S_{c}^{(i)}$, and a SoftMax layer which renders the model outputs in probabilistic terms. }
\label{fig:resnet}
\end{figure}

We construct a ResNet-like \cite{he2015deep} NN, depicted in Fig.~\ref{fig:resnet} to perform the parton-level classification task(s) of our study. 
The model is constructed from dense ResModules which contain a fully-connected (FC) dense layer, followed by an activation layer and a dropout layer with skip connections that alternate layers. We implement four such ResModules with a final output dense layer, for a total of 9 dense layers. The output FC dense layer outputs the classification scores, $S^{(i)}_{c}$, for each example and each class. The final layer is a SoftMax layer which converts the output scores into a vector of probabilities. We employ the ReLU activation function in the ResModules which prevents negative inputs from propagating through the NN activations, while dropout layers are used for regularization to ensure that the ML model does not overfit.

As we are primarily interested in the regions significant to the classification of all PDFs of a given class (per flavor and per phenomenological fit), we must use a scoring method less specific than the gradients corresponding to a single PDF MC replica. However, a global explainability technique would be too broad, as it would not discriminate by member of the error set. For each ratio, we therefore compute standard deviations\footnote{Alternative prescriptions, for instance, based on computing simple averages of the backpropagated classification-score gradients over all members of the test set, produce very similar results.} of the gradients of all PDFs belonging to a specific class at each $x$ point. This allows us to address ML-specific issues that arise from saliency-based interpretations of ML predictions such as those discussed in Refs.~\cite{kindermans2017unreliability,ghorbani2018interpretation,adebayo2020sanity}. The standard deviations tend to be largest in those regions of the PDFs' $x$ dependence corresponding to the most significant shape changes, implying that these regions provide the most sensitive discrimination basis for the PDF classifier. Additionally, we smooth the gradients in $x$-space by evaluating a moving average, further increasing the readability of the saliency maps without compromising their accuracy. Finally, we obtain human-readable saliency maps by plotting the values of our smoothed and normalized standard deviations as colors between the uncertainty bounds of the ensemble of PDF MC replicas.

Along with the saliency maps, we report the confusion matrices for the ML model's predictions of the various PDF classifications. The entries of the matrix are reported as percentages of correct predictions and are normalized across the true labels, of which there are 1500 for each class.

%

\section{\texttt{XAI4PDF} classification results}
\label{sec:framework}
In this section, we summarize our main findings regarding the explainable
classification of PDFs. First, we report the saliency maps and confusion matrices for the parton-flavor classification task within the CT18 NNLO baseline fit. This first study allows us to identify which regions in $x$-space the NN identifies as most important for accurately classifying the parton flavor of the associated PDF. We then perform a similar analysis but with the classification task of discriminating among the CT18(A/X/Z) phenomenological fits. The XAI4PDF method identifies regions of importance that correlate to specific theory choices, highlighting how the NN identifies the manner in which specific choices in the underlying theory change the shape of the downstream fitted PDFs.

\begin{figure}[!ht]
    \centering
    \includegraphics[width=0.6\columnwidth]{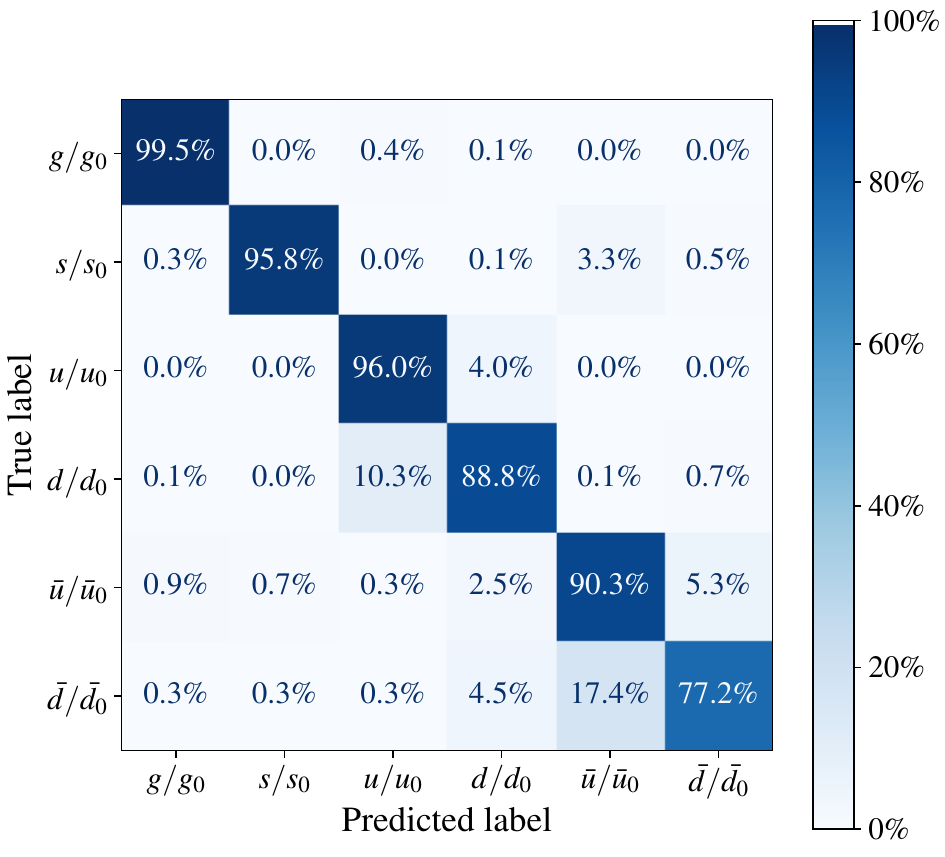}
    \caption{Confusion matrix of the output classification scores for a PDF flavor classifier tested on the CT18 NNLO PDF fit. Depicted are the true PDF ratio labels versus predicted PDF ratio labels, normalized to the true labels, and scores are given in percentages. The color bar indicates scores as well in percentages for presentation clarity.}
    \label{fig:flavor_class_cm}
\end{figure}


\subsection{Parton flavor identification}
\label{sec:pdf_flavor_id}
We first demonstrate XAI techniques within a single phenomenological analysis, CT18 NNLO, by identifying prominent $x$-dependent features informing the accurate classification of the parton flavor associated with each PDF. In particular, the XAI model learns which region(s) in $x$ makes a PDF of specific flavor unique amongst the others. We train the ResNet-like architecture described in Sec.~\ref{sec:class_model} to distinguish among parton flavors by correctly classifying them, specifically looking at $u(x)$, $d(x)$, $s(x)$, $g(x)$, $\bar{u}(x)$, and $\bar{d}(x)$. The model was trained on the ratios of PDF MC error sets to the central fit, with the aim of eliminating the dependence on artifacts of the PDFs' absolute normalizations and instead emphasizing the uncertainty bands. We note that the \texttt{XAI4PDF} approach is of sufficient generality that alternative training and classification strategies based on some admixture
of PDFs' uncertainty bands and absolute [{\it i.e.}, $\sim\!\!f_i(x,Q)$] behavior may also be well-motivated. We defer such detailed studies to future work.

In Fig.~\ref{fig:flavor_class_cm}, we show the results of our trained classification model on the PDF ratio data for the CT18 NNLO fit, reported through a confusion matrix. The model predictions are evaluated on a holdout test set which remains unseen during the training process. The confusion matrix is normalized across rows such that each row sums to $100\%$, with the columns depicting the percent that the predicted label matched the true label. The confusion matrix shows that the ratios for the gluon ($g$/$g_{0}$), strange ($s$/$s_{0}$), and up-quark ($u$/$u_{0}$) PDFs are most accurately classified, with positive identification rates of $99.5\%$, $95.8\%$, and $96\%$, respectively. Conversely, the $d$-quark ratio is confused for that of the $u$-PDF by over $\sim\!10\%$, while the $\bar{d}$ ratio remains unseparated from the $\bar{u}$ ratio by more than $\sim\!17\%$. This confusion between the $\bar{u}$ and $\bar{d}$ ratios is related to statistical indistinguishability arising from flavor-separation challenges in phenomenological fits of the sea-quark densities; this highlights the importance of further experimental measurements of the $\bar{d}$/$\bar{u}$ asymmetry in experiments such as SeaQuest~\cite{Hou:2022sdf,Guzzi:2021fre,Hou:2022ajg}.

\begin{figure*}[!htp]
    \includegraphics[width=\columnwidth]{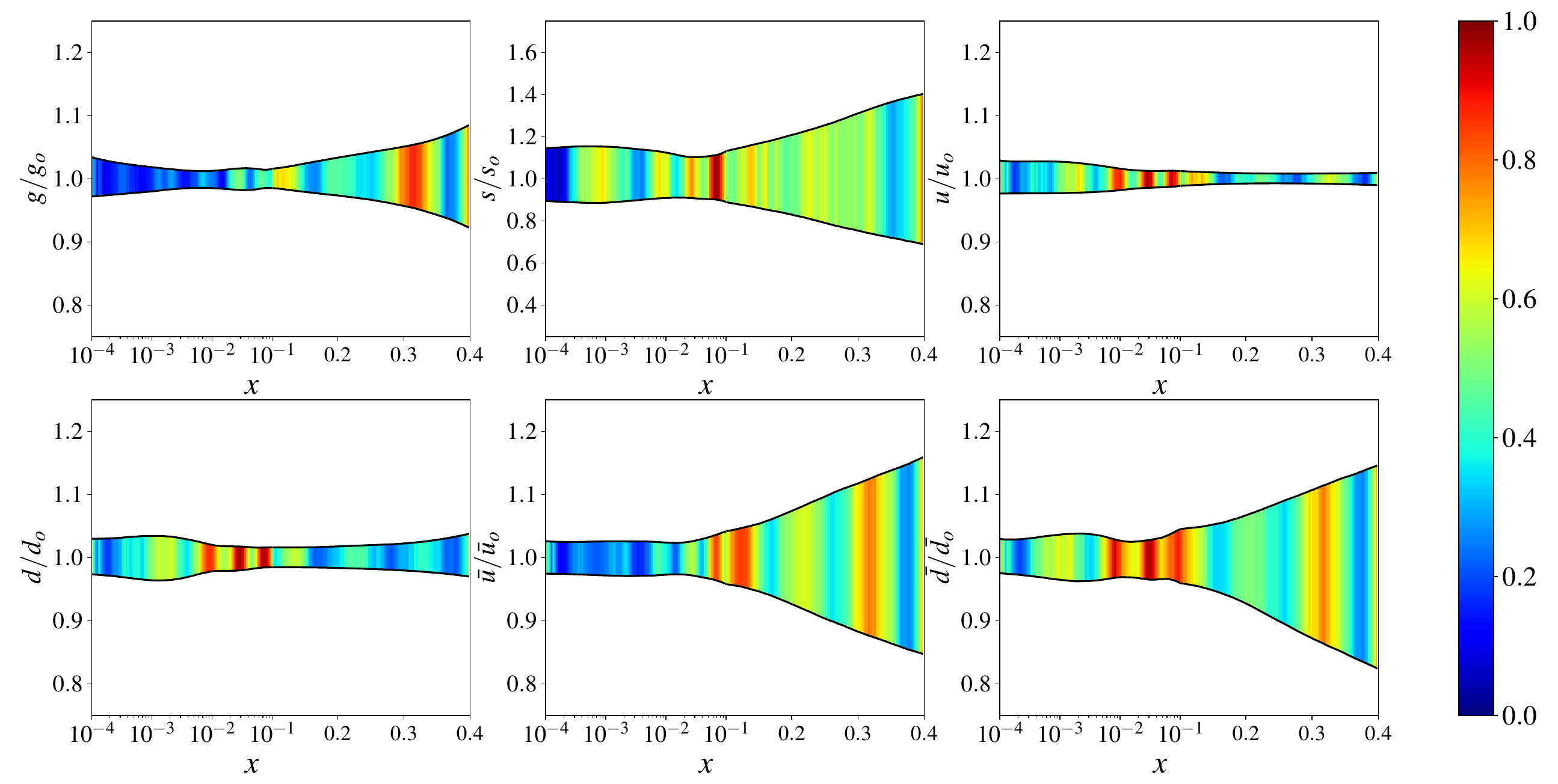}
    \caption{Results of the guided backpropagation algorithm on the trained ResNet-like architecture for PDF flavor classification. The gradients are ensembled and we report a normalized variance. The gradients are superimposed over the uncertainty bands of each PDF at $Q^{2} = 100$ GeV$^{2}$ over which the model was tested. Plotted colors represent the normalized magnitudes of the backpropagated gradient standard deviations as given by the scale at right.}
    \label{fig:flavor_saliency}
\end{figure*}

The results of our guided backpropagation method on the trained ResNet-like architecture for flavor classification are shown in Fig.~\ref{fig:flavor_saliency}. We report the normalized variances for each ensemble of gradients per parton flavor as a color superimposed over the PDF uncertainty band on which the model was tested. We note that each flavor-dependent PDF band has a unique color striping depicting the regions in $x$-space distinguishing that PDF as unique relative to the others. The large variances in the gradients (depicted by red bands) occur in regions where there are more significant shape changes amalgamated over the set of MC replicas as compared to the other PDF flavors. For example, at the $Q\! =\! 100$ GeV scale of Fig.~\ref{fig:flavor_saliency}, the moderate-to-high-$x$ region of $x\! \in\! [10^{-2}\! -\! 10^{-1}]$ is influential for separating the $u$-, $d$-, $\bar{u}$-, and $\bar{d}$-PDFs as indicated by gradient highlighting over that $x$-range. The characteristic region used by the ML model to distinguish the gluon PDF is located at larger $x\! \sim\! 0.3$ (of interest in connection with jet and dijet production~\cite{Hou:2019efy}), while the strange PDF ratios are most discriminated by their $x$ dependence near $x\! =\! 10^{-1}$. 
By visual inspection of the uncertainty band, the regions where the gradients vary the most are regions where there are shape changes that are unique to that particular PDF. We deduce this to be the likely reason the NN model occasionally confuses $\bar{d}$-PDF replicas with those of the $\bar{u}$-PDF, but not reciprocally. The $\bar{d}$-PDF contains replicas which encapsulate more of the features of the $\bar{u}$-PDF replicas than the other way around. This pattern suggests that the $\bar{d}$-PDF may be slightly less constrained than the corresponding $\bar{u}$-PDF for the purposes of the PDF classifier.


\subsection{Disentangling analysis elements among phenomenological fits}
\label{sec:pheno_id}

Many theory settings which enter phenomenological QCD fits can be actively adjusted prior to optimizing the shape parameters
for the PDFs. As argued above, we aim to identify kinematical regions of the PDFs (in particular in $x$-space) where these theory assumptions affect 
the fitted shape(s) of the parton densities. We base our demonstration on the CT18(A/X/Z) analysis series discussed in Sec.~\ref{sec:theory}, which entailed a closed group of systematic variations in the PDF analysis settings suitable for classifier labeling as summarized in Table~\ref{tab:AXZ}.
Much as discussed for the parton-flavor classification task of Sec.~\ref{sec:pdf_flavor_id}, we now train our ResNet-like NN to identify PDF
combinations according to which of the four CT18 fits produced them.

\begin{figure}[!h]
\centering
    \includegraphics[width=0.6\columnwidth]{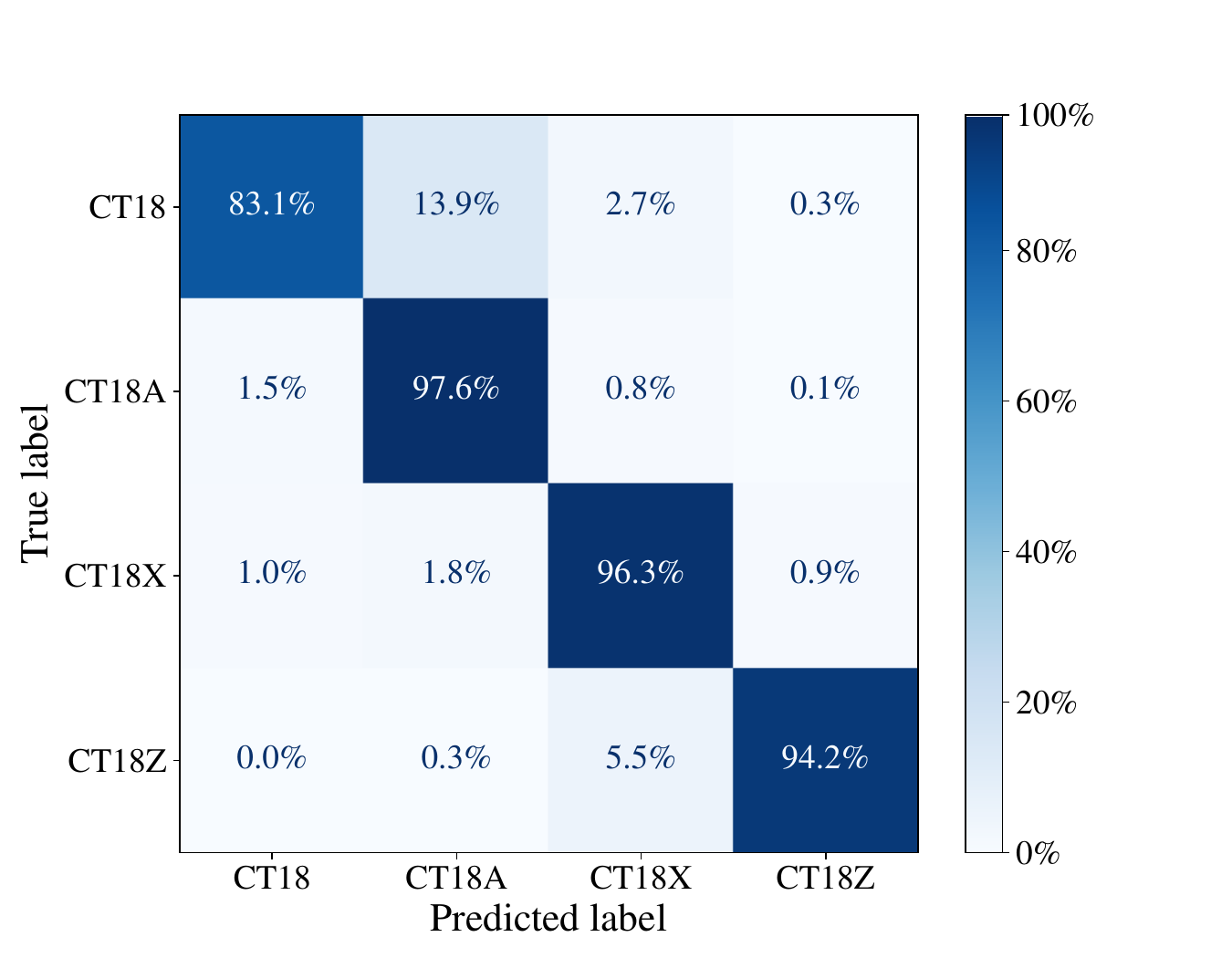}
    \caption{The confusion matrix for the ResNet-like model trained to classify members of the 4-fold collection of CT18(A/X/Z) fits. The scores are given as percentages normalized across the rows according to the true label. The color gradient indicates the score as well for clarity of presentation.}
    \label{fig:pheno_conf}
\end{figure}

In Fig.~\ref{fig:pheno_conf}, we again report the results of our trained ResNet-like model in the form of a confusion matrix based on a holdout test set.
The model generally performs quite well, accurately classifying PDFs belonging to each of the phenomenological fits, with the lowest performance at $83.1\%$. This most common confusion occurs between the CT18 and CT18A replicas; as noted in Table~\ref{tab:AXZ}, the latter of these differs from the CT18 baseline only in the inclusion of the ATLAS 7 TeV data set.
The observed behavior of the confusion matrix is thus consistent with expectation, as the modified DIS factorization scale used in both CT18X and CT18Z has a more apparent impact on the PDF uncertainties over a range of flavors than the inclusion of the ATLAS 7 TeV data in the $x$ regions analyzed~\cite{Hou:2019efy}, especially given that the latter variation primarily affected the strange PDF. Additionally, the CT18Z replicas are confused with CT18X at a rate of $5.5\%$.
This relatively mild rate of confusion likely reflects the fact that, although CT18Z differs from CT18X in every analysis setting toggled
in Table~\ref{tab:AXZ} except the factorization scale, this latter variation of the DIS scale is a dominant effect for the classifier and
somewhat overwhelms the variations related to the ATLAS and CDHSW sets and $m_c$ shift.
It is also interesting to confirm that the two analyses which are ``furthest'' from each other based on the choices of Table~\ref{tab:AXZ}, CT18 {\it vs}.~CT18Z, are also the least confused, confirming that the cumulative shift in analysis assumptions drives the fits' statistical distinguishability as inferred by the XAI calculation.

\begin{figure*}[!htp]
    \centering
    \includegraphics[width=\columnwidth]{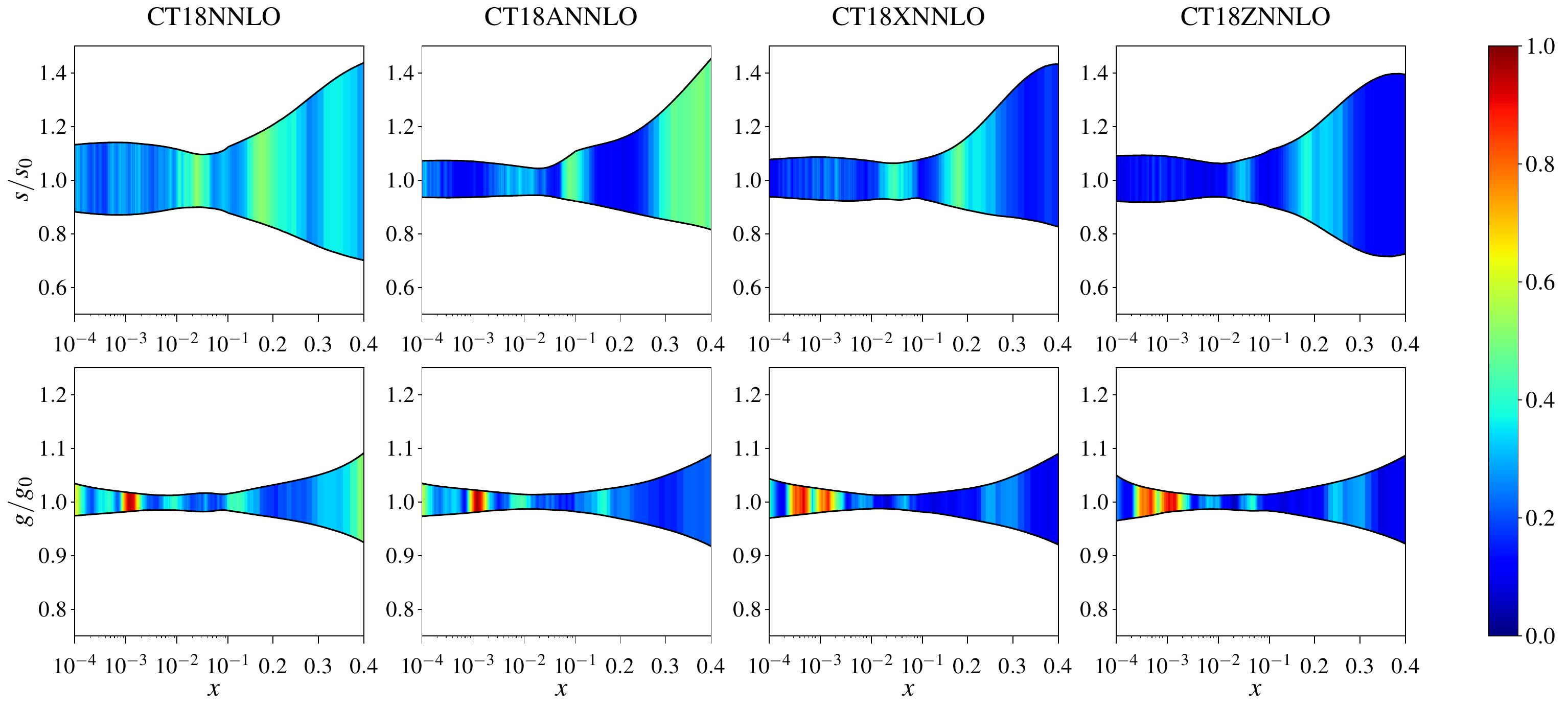}
    \caption{Results of the \texttt{XAI4PDF} guided backpropagation calculation on the trained ResNet-like architecture for classifying across phenomenological fits. The gradients are ensembled and we report a normalized variance. As before, the gradients overlay the uncertainty bands of each PDF at $Q^{2} = 100$ GeV$^{2}$ on which the model was tested.}
    \label{fig:pheno_class_saliency}
\end{figure*}

We show in Fig.~\ref{fig:pheno_class_saliency} the results of the guided backpropagation saliency map applied to the ResNet-like model trained on the PDF replicas from the CT18 series of fits. As in Fig.~\ref{fig:flavor_saliency}, colors indicate variances in the normalized gradients, highlighting regions with prominent feature changes connected to the ML model's fit classifications. In general, the model identified the gluon- and strange-PDF ratios as being most significant in distinguishing among the phenomenological fits, as evidenced by the gradient patterns. In contrast, the other PDF ratios which we have not explicitly plotted ({\it i.e.}, the scored ratios for $u$, $d$, $\bar{u}$, $\bar{d}$) showed very little gradient structure, indicating that the model did not identify significant feature changes among the various fits for these other PDF replicas. Consequently, we report only the gluon and strange saliency maps as corresponding to the PDF replicas most important in the guided backpropagation model classifications. We find this observation to align closely with the findings in Ref.~\cite{Hou:2019efy}, which explored the significant PDF impacts of the CT18 fit variations with special attention on the gluon and strange parton densities. The ability of the model to pick out this behavior and emphasize these variations suggests that it is effectively learning the flavor-dependent effects of the varying theory assumptions in the fits.

The results for the strange- and gluon-PDF ratios shown in Fig.~\ref{fig:pheno_class_saliency} demonstrate additional interesting behavior.
For example, it is noteworthy that the gluon replicas play a dominant role in the classifications among the CT18 series, as reflected by the large peaked values of the gradient deviations in the neighborhood of $x\!\sim\!10^{-3}$. The strange-PDF replicas have smoother, less peaked gradients, indicating that their relevance in the classification is important but perhaps not as strong for the ML model. The gradients tend to peak in some cases at $x\!\sim\!10^{-2}$ but also at more moderate-$x$ values such as $x\!\sim\!0.2$. The $x$-regions which are highlighted by the gradients are those where the uncertainty changes most between fits. These results are supported by Ref.~\cite{Hou:2019efy}, where the $s$ and $g$ PDF uncertainties are found to be affected more dramatically in those regions, particularly at low-$x$.

Finally, it is natural to ask whether meaningful correlations exist which might further reinforce the interpretability
of the classification model, {\it e.g.}, between the NN classification and initial PDF; such quantities could
yield additional insights into the meaning of the saliency maps, while also possessing calculable analogues
in the input feature spaces on which the classification is trained.
Borrowing an analogy to Hessian analyses, we note that the $L_2$ sensitivity method~\cite{Hobbs:2019gob,Jing:2023isu} can be defined
in the parametric vicinity of the global $\chi^2$ minimum. This leads to the definition, 
\begin{equation}
	S_f = \vec{\nabla} \chi^2_E \cdot {\vec{\nabla} f \over |\vec{\nabla} f|} = \Delta[\chi^2]\ \mathrm{Corr}_\mathrm{H}(\chi^2,f)\ .
\end{equation}
On the logic that classification scores can be computed on the basis of MC replicas generated from an original Hessian
analysis and informed by notions of statistical separability in that progenitor analysis, an approximate analogue expression
for the pre-SoftMax classification score, $S_c$, may assume the form
\begin{equation}
	\Sigma_c \equiv \vec{\nabla} S_c \cdot {\vec{\nabla} f \over |\vec{\nabla} f|} \sim \Delta[S_c]\ \mathrm{Corr}_\mathrm{MC}(S_c,f)\ ;
\end{equation}
this would be reasonable for scenarios in which classifications can be directly projected onto the Hessian fits by which they were
produced.
Ultimately, developing a more systematic matching between ML-based classification methods like those embodied by the
\texttt{XAI4PDF} framework could thus proceed on the basis of such statistically-matched calculations, an
undertaking we expect to pursue in future work.

%

\section{Conclusion}
\label{sec:conc}
In this study, we have explored for the first time NN explainability techniques for PDF phenomenology through a novel framework, \texttt{XAI4PDF}, applied to the task of PDF classification and model discrimination. Using this approach, we classified PDF replicas within a particular phenomenological fit by flavor and identified key regions in the $x$-space that distinguished them. Additionally, we classified PDF replicas across phenomenological fits with incrementally shifted analysis internals, uncovering specific $x$-space regions corresponding to these changes. We calculated confusion matrices for each fit to ensure that our models achieved peak accuracy, and to understand the specific errors made during classification and how those errors related to aspects of the phenomenological fit.

We conclude with thoughts on the generalizability
of the calculations explored above as they relate to full-scale fitting frameworks.
With sufficient sampling, it may be possible
to discriminate the downstream realization of very subtle
theoretical corrections or effects within a full phenomenological fit, including those related
to possible BSM scenarios. The use of the guided backpropagation technique with its double masking procedure during autodifferentiation allows for highly expressive and fine-grained gradient structures, as one would expect from potential BSM contaminations.

In this work, we utilized a single explainability technique in the form of guided backpropagation that fit the parameters of our specific problem; however, many other explainability techniques exist that can be employed for different but complementary problems in PDF theory.
A further opportunity for improvement pertains to developing additional quantitative checks of the method~\cite{2023NatSR..1316887S} against
which to correlate the saliency maps, for instance, along the lines of the discussion at the end of Sec.~\ref{sec:pheno_id}.
Extending \texttt{XAI4PDF} with quantifiable measures of uncertainty would allow for a deeper and more comprehensive analysis of the impact of variations in the PDF fitting procedure. 

This approach could also aid in the construction of PDF parametrizations by visualizing the impact of different theoretical choices implemented during fitting. Explainability techniques, as demonstrated in this paper, could serve as valuable tools to analyze how and where varying the PDF parameters affects the downstream uncertainty. A goal that is particularly attractive for the advancement of high-precision SM tests and new physics discovery at future machines where uncertainty quantification is crucial. By leveraging novel ML explainability techniques, this work suggests options for more precise studies of parton densities.
Moreover, refinements of the approach used in this study would be further applicable within analyses of the type
leading to the combined fits of the PDF4LHC variety, which are based on constructing MC samples from a series of phenomenological
analyses before statistically enfolding these into a single amalgamated set.
The \texttt{XAI4PDF} method would permit an investigation of the statistical compatibility of input sets in the apples-to-apples
MC basis.

\vspace{0.3cm}

{\bf Code availability}. We release a working version of \texttt{XAI4PDF} for example cases of both PDF flavor and phenomenological fit classification at
a dedicated GitHub repository, \url{https://github.com/ML4HEP-Theory/XAI4PDF}.

\section{Acknowledgments}
The work of BK and TJH at Argonne National Laboratory was supported by the U.S.~Department of Energy under contract DE-AC02-06CH11357.
The work of JG was supported by the U.S.~Department of Energy, Office of Science, Office of Workforce Development for Teachers and Scientists (WDTS) under the Science Undergraduate Laboratory Internship (SULI) program.
%
%

%

\bibliographystyle{utphys}
\bibliography{xai}

\end{document}